\documentclass{aastex62}
\usepackage{graphicx}

\usepackage{amsmath}

\received{July 25, 2019}
\revised{\today}
\accepted{ }
\submitjournal{ApJ}

\shorttitle{Source Position and Duration of Solar Type III Burst}
\shortauthors{Zhang et al.}

\begin{document}
	
	\title{On the Source Position and Duration of a Solar Type III Radio Burst Observed by LOFAR}

	\correspondingauthor{ChuanBing Wang}
	\email{cbwang@ustc.edu.cn}
	
	\author[0000-0001-6855-5799]{PeiJin Zhang}
	\affil{CAS Key Laboratory of Geospace Environment,
		School of Earth and Space Sciences, \\
		University of Science and Technology of China,
		Hefei, Anhui 230026, China}
	\affiliation{CAS Center for the Excellence in Comparative Planetology, Hefei, Anhui 230026, China}
	
	\author[0000-0003-2872-2614]{SiJie Yu}
	\affiliation{New Jersey Institute of Technology, Newark, NJ, 07102, USA}

	\author[0000-0002-8078-0902]{Eduard P. Kontar}
	\affiliation{School of Physics and Astronomy, University of Glasgow, Glasgow G12 8QQ, UK}
	
	\author[0000-0001-6252-5580]{ChuanBing Wang}
	\affiliation{CAS Key Laboratory of Geospace Environment,
		School of Earth and Space Sciences, \\
		University of Science and Technology of China,
		Hefei, Anhui 230026, China}
	\affiliation{CAS Center for the Excellence in Comparative Planetology, Hefei, Anhui 230026, China}
	
	
	
	\begin{abstract}
		The flux of solar type III radio burst has a time profile of rising and decay phase at a given frequency, which has been actively studied since 1970s. Several factors that may influence the duration of a type III radio burst has been proposed. In this work, to study the dominant cause of the duration, we investigate the source positions of the front edge, the peak, and the tail edge in the dynamic spectrum of a single and clear type III radio burst. {The duration of this type III burst at a given frequency is about 3 second for decameter wave}. The beam-formed observations by the LOw-Frequency ARray (LOFAR) are used, which can provide the radio source positions and the dynamic spectra at the same time. We find that, for this burst, the source positions of the front edge, the peak, and the tail edge split with each other spatially. The radial speed of  {the electrons exciting} the front edge, the peak, and the tail edge is 0.42\,c, 0.25\,c, and 0.16\,c, respectively. We estimate the influences of the corona density fluctuation and the electron-velocity dispersion on the duration, and the scattering effect by comparison with a few short-duration bursts from the same region. The analysis yields that, in the frequency range of 30\,--\,41\,MHz, the electron-velocity dispersion is the dominant factor that determines the time duration of type III radio bursts with long duration, while scattering may play important role on the duration of short bursts.
	\end{abstract}
	
	
	
	\section{Introduction}
	
	Solar type III radio burst is generated by electron beams injected into the open magnetic field line  {in the corona}. These electron beams passing through tenuous plasma can generate Langmuir waves by triggering the bump-in-tail instability. Consequently, part of the electrostatic Langmuir waves may convert to electromagnetic (EM) waves through non-linear process \citep{ginzburg1958, melrose1980, reid2014review}. This is the plasma emission mechanism of type III radio burst. The EM waves propagate through the disturbed refractive corona and relatively transient interplanetary space before received by the observer. The outward speed of the type III radio burst exciter is about 0.1\,--\,0.5 c in the solar corona (e.g. \citealp{dulk1987speeds}; \citealp{reiner2015electron};  \citealp{zhang2018type}), where c is the speed of light in vacuum. The exciter moves through the lower corona within tens of seconds. This time scale is much shorter than the large-scale density variation in the corona. Thus, the type III radio burst can be used to diagnose the density structure of the corona.
	
	The type III radio bursts are normally intense and observed over wide range of frequencies. With high time resolution radio spectrometer, the rising and decay phase of the flux intensity (time profile) in a given frequency channel can be observed. Based on the plasma emission mechanism, there are several factors that may contribute to the time duration of type III radio burst:
	\begin{itemize}

		\item	The velocity dispersion of the electron beam exciter (also called free-stream effect). The electron beams generated at the acceleration region may have different speed. The faster electrons arrive earlier at the height of the radio-wave source region, while the slower electrons arrive later. Thus, for a given frequency, the waves are generated in a time period between the time when the fast and slow electrons passing the corresponding height respectively. This thought was recently discussed by \cite{reid2018solar}. They statistically studied the time characteristics of 31 type III radio bursts using the dynamic spectra observed by LOw-Frequency ARray (LOFAR, \citealp{van2013lofar}). Their work stated that the beam elongation caused by the beam spatial expansion is essential for the duration of the type III radio burst.
		\item	The density fluctuations in the corona. The fluctuation of corona density can lead to the fluctuation of the source height of radio waves with the same frequency, therefore these waves are generated at different time. This idea is proposed by \cite{roelof1989type}. According to their calculation,  {the electron density variation of $\left\langle \delta N_e/N_e\right\rangle \simeq 0.3$} at a given height alone can produce the time profile for meter and decameter type III radio burst.
		\item	The wave propagation effect. Waves generated at the same time from the same position may arrive at the observer at different time due to scattering and refraction. Previous studies \citep{fokker1967coronal, steinberg1971coronal, thejappa2007monte} used radio ray-tracing method to simulate the wave propagation effect. The simulations indicate that the scattering and the refraction can generate the fast rise and exponential decay phase from an ideal ‘pulse’ in the corona.
		\item	The intrinsic emission process of the radio wave. The electromagnetic wave of the type III radio burst has an intrinsic growth and damping process.
	\end{itemize}
	In general, the duration or time profile of type III radio burst is a convoluted result of the wave excitation, the extended source, and the wave propagation process. Simulations on the generation, and propagation of type III radio burst with different assumptions \citep{li2008simulations1, li2008simulations2, li2009simulations3,kontar2009onsets,reid2015stopping} can get a duration which is comparable with the observed value.
	
	However, it is difficult to {observationally determine} the dominant cause of the duration just from the dynamic spectrum. The observation of the positions of the radio source at different time and frequency, accompanying with the dynamic spectrum, is in need. LOFAR is an advanced radio antenna array with the capability of producing the dynamic spectra and radio image at the same time \citep{van2013lofar}. It has two kinds of array observing in two frequency bands, the Low Band Array (LBA) in the frequency range of 10\,--\,90 MHz and High Band Array (HBA) in the frequency range of 110\,--\,250 MHz. LOFAR has 50 stations, 38 of them are located in the North-East of Netherlands, and 12 international stations are located in Germany, Poland, France, Sweden and UK. It is capable of a variety of processing operations including correlation for standard interferometric imaging, the tied-array beam-forming, and the real-time triggering on incoming station data-streams. The beam-forming mode can produce the dynamic spectra and source imaging data of the radio burst with high temporal and frequency resolution at the same time. With this dynamic spectrum imaging data, we can investigate the relationship between the source positions and the duration by analyzing the position of the radio source at different time and frequency.
	
	In this work, we intend to study the source positions and the factors that may influence the duration of a type III radio burst observed by LOFAR on 6 May 2015. The paper is arranged as follows, in Section 2, we introduce the background of this event. In Section 3, we display the radio imaging spectroscopy of this single, clear and smooth type III radio burst. Section 4 contains the analysis for the factors that contribute to the duration of the type III radio burst. Section 5 is the summary.
	
	\section{The type III burst on 6 May 2015}
	
\begin{figure}[h]
		\centering
		\includegraphics[width=0.9\textwidth]{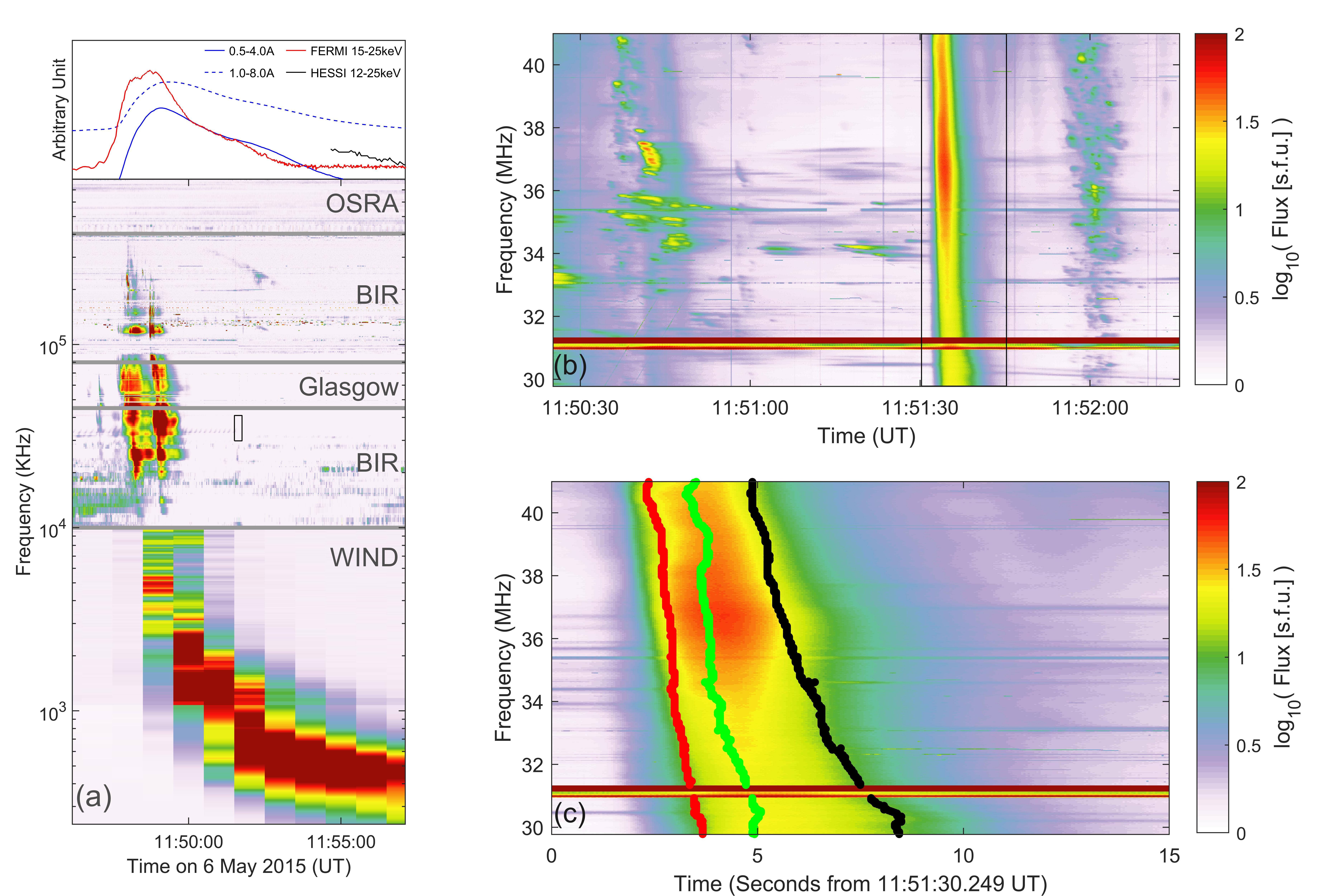}
		\caption{The dynamic spectra of the type III bursts on 06 May 2015. Panel (a) shows the overview of the type III radio burst group and the {X-ray flux from GOES \citep{menzel1994introducing}, FERMI \citep{meegan2009fermi} and RHESSI \citep{lin2003reuven}}. The dynamic spectrum is a joint observation from {e-Callisto stations (OSRA, BIR, Glasgow)  and space-based (WIND)} radio observation. Panel (b) is the dynamic spectra observed by LOFAR. The time and frequency range of Panel (b) is marked by the black box in Panel (a). Panel (c) is the dynamic spectrum of the type III radio burst studied in detail in this work, where the red, green, and black solid lines are the front edge, the peak, and the tail edge of the burst  {with definitions in the main text section 3}. The starting and end time of Panel (c) are indicated by two black vertical lines in Panel (b). 	\label{fig:1}}
	\end{figure}

	The radio burst that we study took place at about 11:51:30 UT on 06 May 2015. It was a single and clear type III radio burst (shown in Figure \ref{fig:1}(c)) with duration of about 3 seconds in the frequency range of 30\,--\,41\,MHz from the LOFAR observation. It happens after a strong branch of type III radio burst group (shown in Figure \ref{fig:1}(a)). The type III burst group started from 11:47\,UT and covered from 0.1 to 400\,MHz in dynamic spectrum of the joint observation by {e-Callisto \citep{benz2009world}} and {WIND \citep{lin1995three}}. Unfortunately, the main part of this type III group between 11:47\,UT and 11:50\,UT is saturated in LOFAR observation. In this work, we will focus on the single and clear type III radio burst after the type III group starting at 11:51:30 UT (marked with black box in Figure \ref{fig:1}(b) and shown in Figure \ref{fig:1}(c) in detail). The beam-formed data of LOFAR is used to study the relationship between the duration and the source positions of this type III radio burst.  {The temporal and frequency resolution of the data are 52.4 ms and 24.2 kHz, respectively.}
	
\begin{figure}[h]
		\centering
		\includegraphics[width=0.9\textwidth]{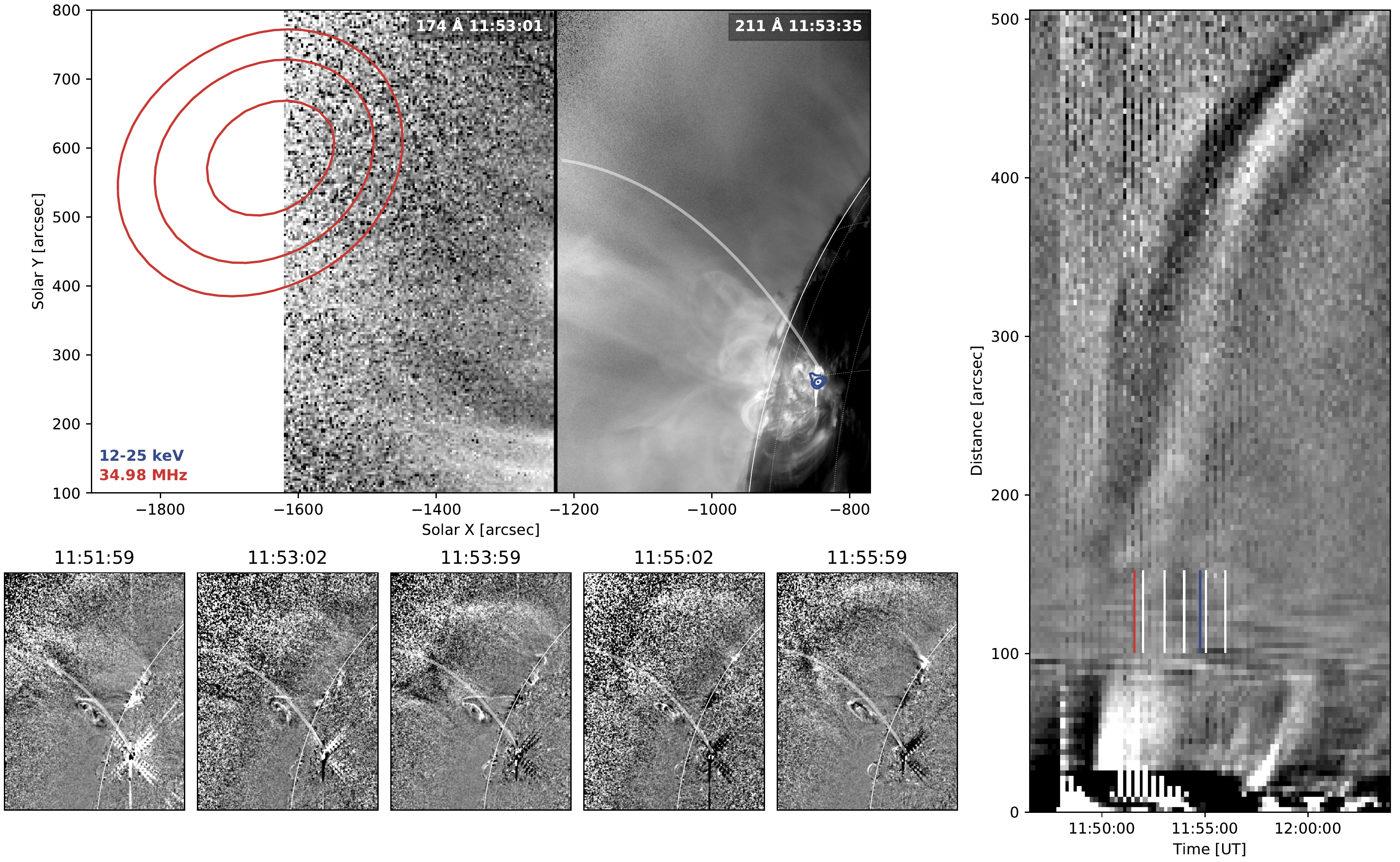}
		\caption{The overview of the position relationship of the flare and the radio source observed by LOFAR, RHESSI HXR \citep{lin2003reuven}, SDO/AIA, and the Proba2/SWAP \citep{berghmans2006swap}. The left upper panel shows the AIA 211 \AA{} image and the running difference of SWAP 174 \AA{} overlapped with the hard X-ray image  (blue contour) and the radio source (red contour), The corresponding times of the radio image and the Hard X-Ray image are marked as right and blue bars in the right panel, the gray line indicates the inferred magnetic field line connecting the radio source and the flare site. The five running difference snapshots in the lower-left panel shows the evolution of the small flare associated with the type III radio burst. The small flare exploded at 11:51:11\,UT according to the SDO/AIA image. The corresponding times of the snapshots are marked as white bars in the right panel. The right panel shows the time-distance plot along the gray line marked in the left upper panel.\label{fig:2}}
	\end{figure}

	At the time of these radio bursts, there are a few active regions (ARs) on the solar disk. Within these ARs, there is only one AR (AR12339) in its active phase according to the dynamic extreme ultraviolet (EUV) image of SDO/AIA \citep{SDOAIA}. In this AR, a M1.9 flare occurred at 11:48\,UT, {followed with a small flare at 11:51\,UT. The strong type III burst group are produced by the big flare, as can be seen from the good temporal correlation between the radio dynamic spectrum and the X-ray flux in Figure \ref{fig:1}(a). From the temporal and spatial relationship between the radio source and the EUV, hard X-ray (HXR) image shown in Figure \ref{fig:2}, we believe that, for the single type III bust interested, the electron beam of the radio source comes from the small flare}. {For the HXR image reconstruction in Figure \ref{fig:2}, we use the energy band of 12\,-\,25\,keV, the integration time is 4 seconds, the selected detectors are 4F, 5F, 6F, 7F, 8F.} According to the HXR image, the position of HXR flux peak is located at N15E67 at 11:54\,UT. In the following study, we will use this position as the original source position of the electron beams exciting the radio burst. This event is located at 0.93 solar radius from the solar center on the solar disk. As a result, the projection effect can be ignored, and the height of the radio source can be obtained directly from the two-dimensional (2-D) radio image in the sky plane.

	\section{Method of Measurement and Results} 
	
	We use the tied-array beam observation of LOFAR. Its high temporal and frequency resolution and the ability of imaging enables us to study the spectroscopy and the imaging characteristics of the radio source at the same time.
	
\begin{figure}[h]
		\centering
		\includegraphics[width=0.7\textwidth]{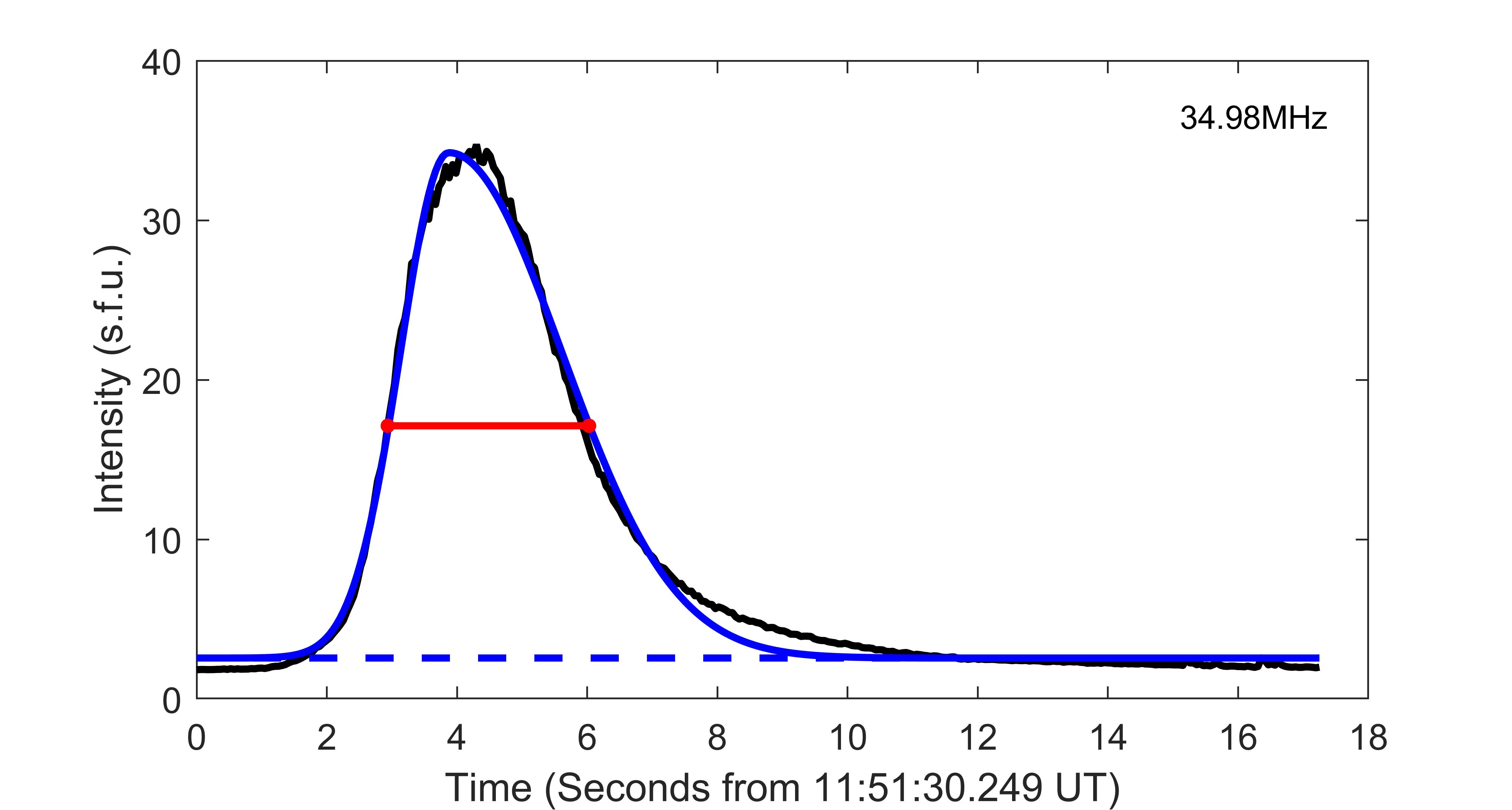}
		\caption{The flux intensity profile of the type III radio burst in the frequency channel 34.98 MHz. The black line is the flux observed by LOFAR. The blue line shows the fitted curve, and the red line-segment marks the FWHM. 
			\label{fig:3}}
	\end{figure}

	To get the observed duration. We use an asymmetric exponential function to fit the light curve of the type III radio burst  {at a given frequency channel}, as suggested by \cite{reid2018solar}, which can be expressed as
	
	\begin{equation}
	I(t) = A \exp{\left(-\frac{(t-t_0)^2}{2 \tau^2}\right)}, \qquad \tau = \left\{ \begin{aligned}
	\tau_{R}, & & {t \leq t_0}\\
	\tau_{D}, & & {t >  t_0}
	\end{aligned}\right. ,
	\end{equation}
	where $I$ is flux intensity, $t_0$ is the time of peak flux,  $\tau_R$ and $\tau_D$  are two parameters characterizing the rise and decay time respectively. The duration is defined as the full width of half maximum (FWHM) of $I(t)$, which is expressed as $\Delta t_{obs}^\mathrm{FWHM} =(\tau_{R}+\tau_D)\sqrt{2 \ln2}$. As an example, Figure \ref{fig:3} shows the asymmetric Gaussian fitting of the light curve in the frequency channel of 34.98\,MHz, where the duration or FWHM is marked as a red solid line. From Figure \ref{fig:3}, one can see that the asymmetric {Gaussian} function fits well to the flux profile, and the duration obtained is 3.09 second. The sixth column of Table (\ref{tab:1}) lists the duration obtained by this method in six frequency channels  {between 34 and 38\,MHz} with the frequency shown in the second column.

	In this work, we mainly focus on radio waves with intensity above the half maximum in the light curve. The front time $t_1$ and tail time $t_2$ is defined as the time when the fitted flux equals to half of the peak flux on the rising side and decay side  {of the light curve at a given frequency}, respectively. Applying this method to each frequency channel, we can obtain the front edge, the peak, and the tail edge of the burst in the dynamic spectrum. The front edge, the peak line and the tail edge are the frequency-time curve connecting the front time, the peak time and the tail time in all frequency channels  {in the dynamic spectrum}, which are marked as green, red and black lines in Figure \ref{fig:1}(c), respectively.
	
\begin{figure}[h]
		\centering
		\includegraphics[width=0.6\textwidth]{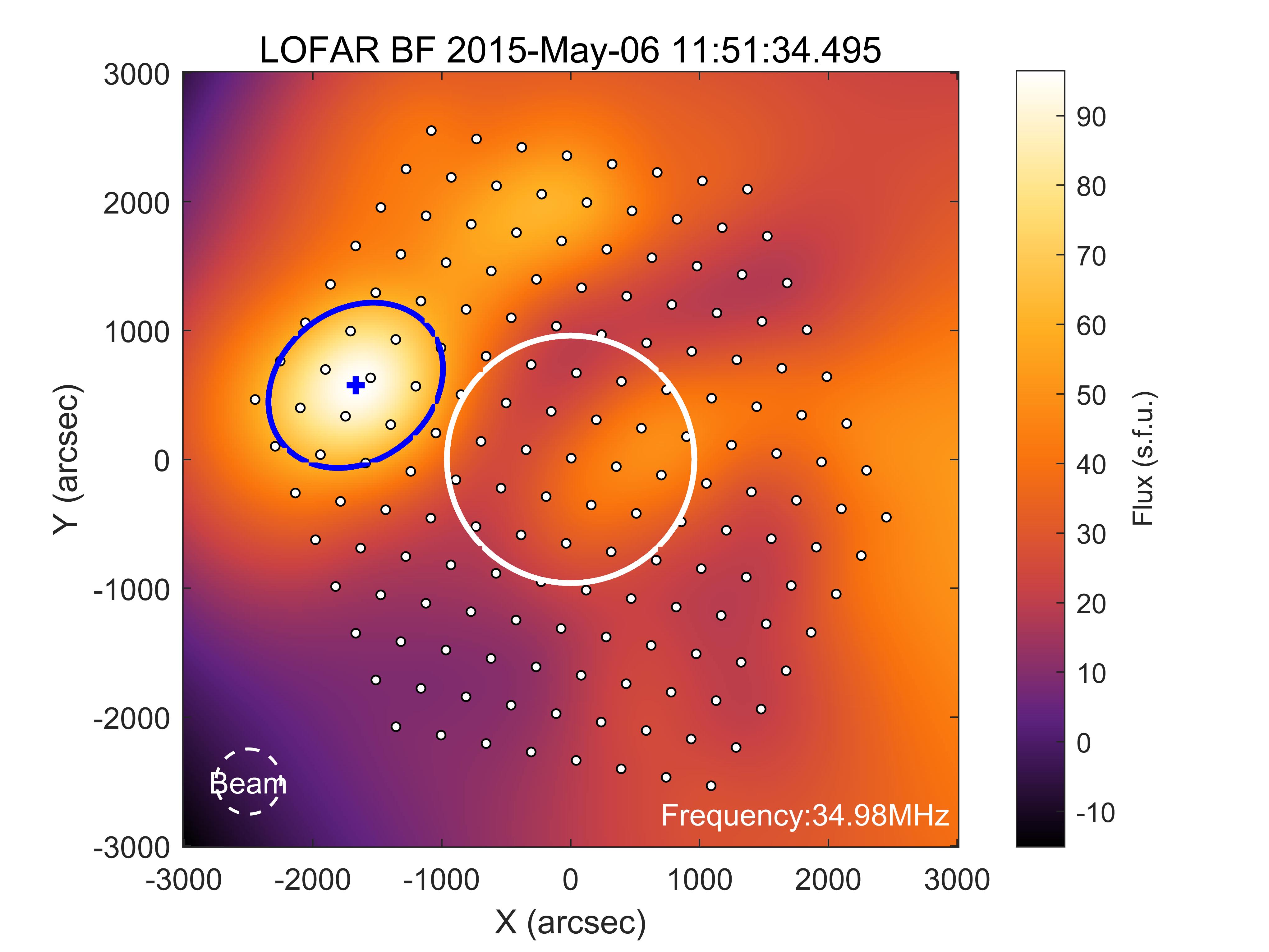}
		\caption{LOFAR Beam formed image of the radio burst at 34.98 MHz and 11:51:34.459\,UT. The blue eclipse is the FWHM of the fitted radio flux distribution. The centroid position is marked with the blue ‘plus’. The white solid circle marks the solar disk. The white dots mark the positions of the beam pointing directions of LOFAR. The LOFAR beam size at the considered frequency is shown by white dashed circle in the left-lower corner. 	\label{fig:4}}
	\end{figure}

	To obtain the size and central position of the radio source  {at a given time and frequency}. We fit the  {flux intensity of} beam-formed radio image using an elliptical Gaussian function \citep{kontar2017imaging}, which is defined as
	\begin{equation}
	S(x,y) = S_0 \exp{\left(-\frac{x'^2}{\sigma_x^2} -\frac{y'^2}{\sigma_y^2} \right)},
	\label{eq:2}
	\end{equation}
	and
	\begin{align}
	x' = (x-x_s)\cos(T)-(y-y_s)\sin(T), \\
	y' = (x-x_s)\sin(T)+(y-y_s)\cos(T), 
	\end{align}
	where $S_0$ is the peak flux of the source, $(x_s, y_s)$ are the coordinates of the central source position, $\sigma_x$ and $\sigma_y$ are the root mean square (RMS) lengths, $T$ is the rotation angle from the x-axis. The fitting method is to minimize the target function
	\begin{equation}
	\chi^2 = \sum_{i=1}^N \frac{(F_i-S(x_i^b,y_i^b;S_0,x_s,y_s,\sigma_x,\sigma_y,T))^2}{\delta F^2},
	\end{equation}
	where $F_i$ is the value of flux observed by the $i$th beam pointing to the position $(x_i^b, y_i^b)$, $\delta F$ is the uncertainty of the flux (usually 1 s.f.u.). Figure \ref{fig:4} shows the LOFAR radio image and the source fitting results for the radio waves in the frequency channel of 34.98\,MHz at 11:51:34.459\,UT on 6 may 2015.
	
	Using this method, we are able to obtain the source size and centroid positions of the type III radio burst at any time-frequency point in the dynamic spectrum. For the convenience of discussion, we define a time ratio as
	\begin{equation}
	t_r  =  \frac{t - t_1}{t_2-t_1},
	\end{equation}
	where $t$, $t_1$ and $t_2$ are the observation time, the front time and the tail time in UT in the light curve, respectively. Thus, $t_r$ can be recognized as a ‘normalized time’. The values of $t_r=0$ and $t_r=1$ correspond to the points on the front edge and the tail edge in the dynamic spectrum, respectively.
	
	\begin{figure}[h]
		\centering
		\includegraphics[width=0.6\textwidth]{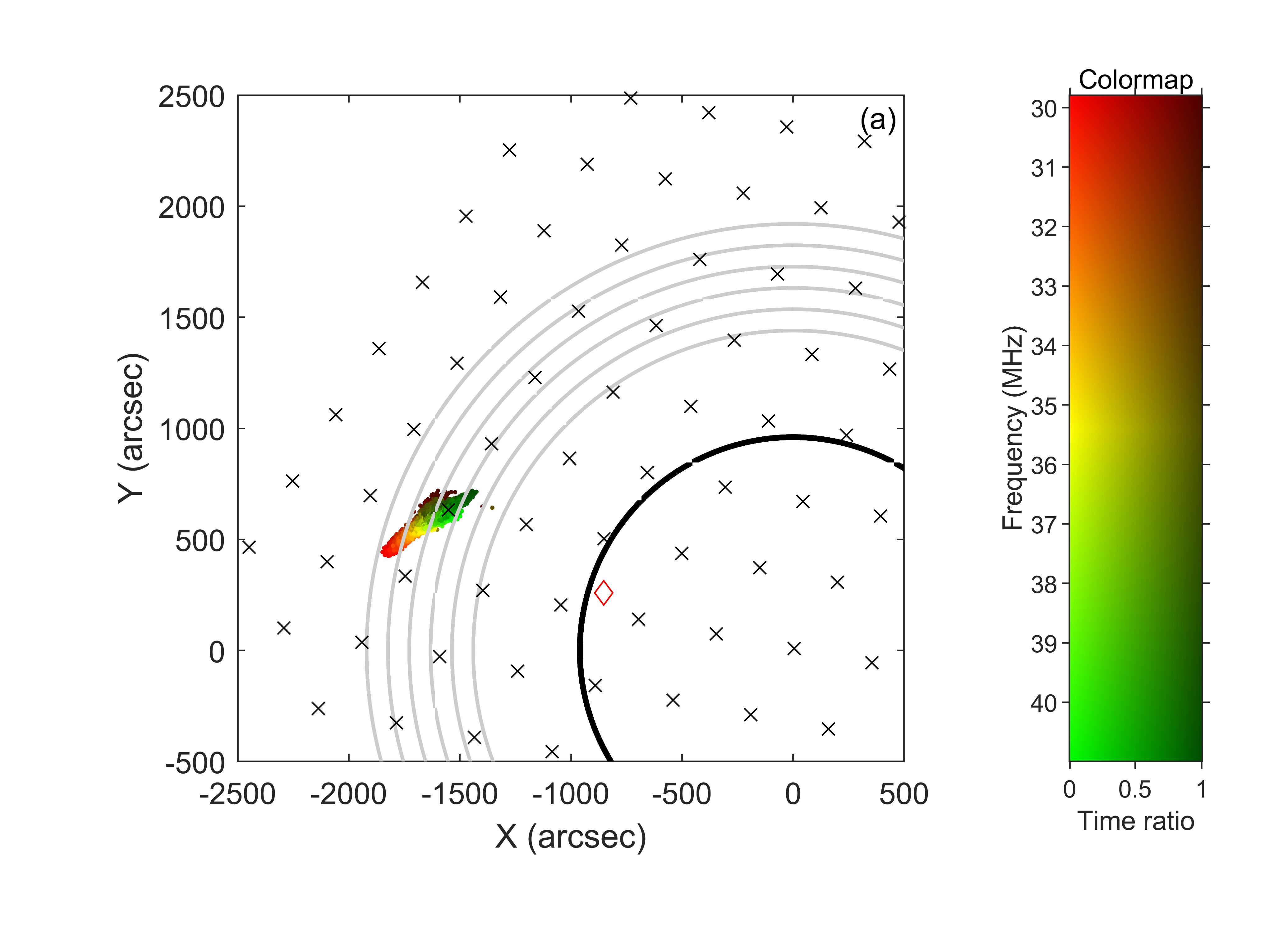}
		\includegraphics[width=0.6\textwidth]{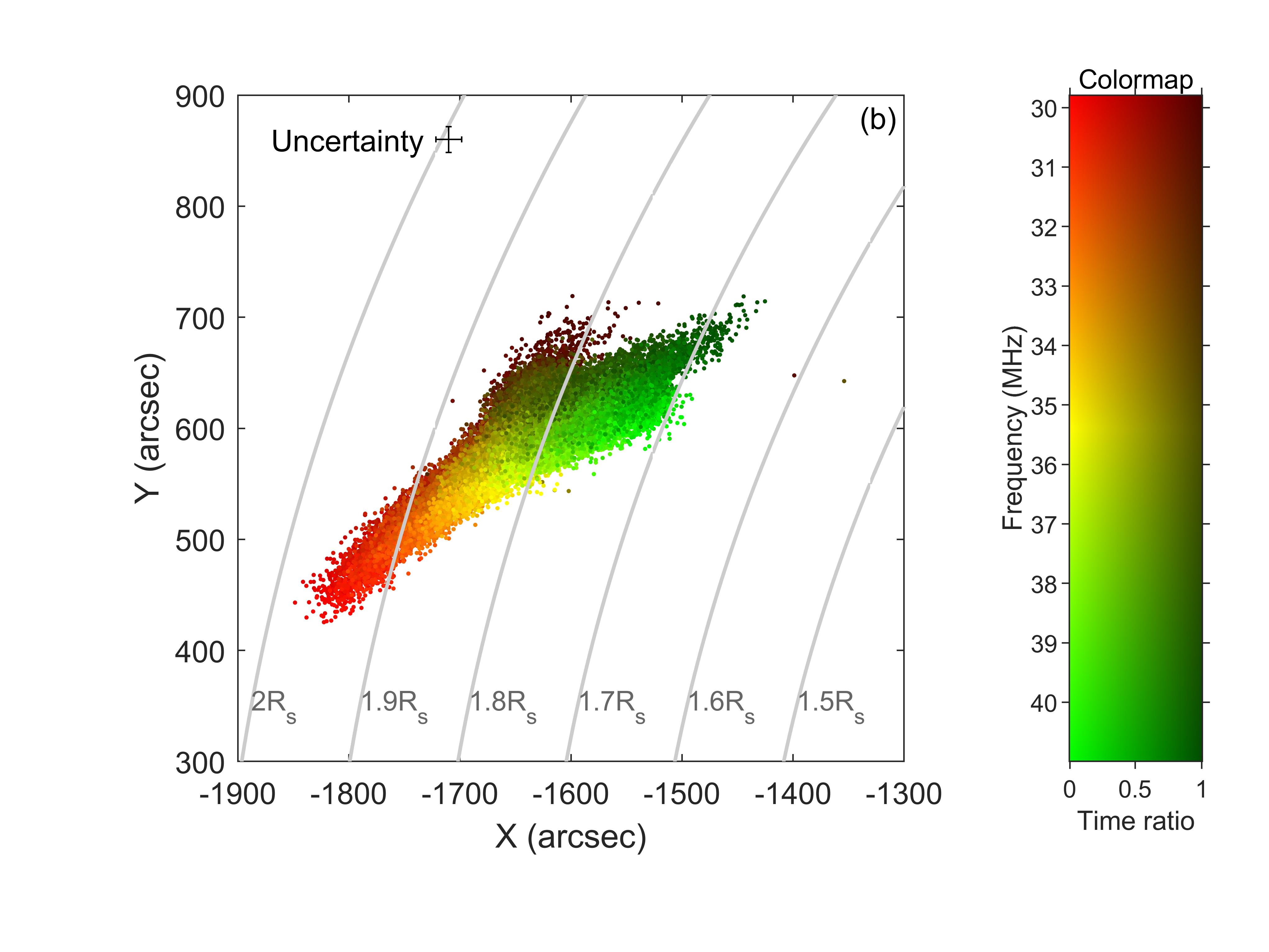}
		\caption{Centroid locations of the sources at different time ratio and frequency for the type III radio burst shown in Figure \ref{fig:1}(c). The points are colored according to the frequency $(f)$ and the time ratio $(t_r)$. The hue of the color represents the frequency of the source, warmer color marks the higher frequency. The saturation of the color marks the time ratio of the source, more chromatic colored points represent smaller $t_r$ (close to the front edge), and \textit{vice versa}.  {The values of $t_r=0$ and $t_r=1$ represent the front edge and the tail edge in the dynamic spectrum, respectively.} Panel (b) is the detailed zoom-in of Panel (a). The average positioning uncertainty is marked as error bar in Panel (b). The flare site is marked by a diamond in Panel (a). \label{fig:5}}
	\end{figure}
	
	Figure \ref{fig:5} shows the source centroid position at different time ratio and frequency channels for the type III radio burst that is shown in Figure \ref{fig:1}(c). The average uncertainty of the source positioning is marked as an error bar in the legend of Figure \ref{fig:5}(b). The uncertainty is given by (see appendix in \citealp{kontar2017imaging})
	\begin{equation}
	\delta x_s \approx \sqrt{\frac{\pi}{2}} \frac{\sigma_x}{\sigma_y}\frac{\delta F}{S_0} h,  \,\,\,\, 
	\delta y_s \approx \sqrt{\frac{\pi}{2}} \frac{\sigma_y}{\sigma_x}\frac{\delta F}{S_0} h,
	\end{equation}
	where $h$ is the angular resolution, which is equal to the beam width. The average positioning uncertainty for this event is about 0.19 arcminute.

	From Figure \ref{fig:5} we can see that, the sources of this type III radio burst are located out of the optical solar disk limb and on a line which passes through the solar disk center and the flare site in the sky plane. This is consistent with that this radio burst originates from a flare site close to the limb. In the following analysis, we assume the projection effect can be neglected when we calculate the height (or heliocentric distance) of the source centroid from the observed position.

	\section{Factors Contribute to the Duration}
	With the centroid positions of the radio source at different time and frequency, we discuss the contribution of different factors to the duration of this type III burst in this section.
	
	\subsection{Effect of Electron-Velocity Dispersion}
	
	From Figure \ref{fig:5}, one can find two qualitative trends on the variation of the radio source position  {with time and frequency}. Firstly, the source height generally increases with the decrease of wave frequency. This is likely to be due to the outward movement of the electron beam exciter and the decrease of the corona density with the increase of altitude. Secondly, for radio waves with the same frequency but arriving the telescope at different time, their source positions displace gradually with time in the north-west direction. Moreover, the later the arrival time is, the lower the source altitude is. This seems to indicate that the electron beams exciting these radio waves are moving along different magnetic field flux tubes with different speed.
	
	\begin{figure}[h]
		\centering
		\includegraphics[width=0.44\textwidth]{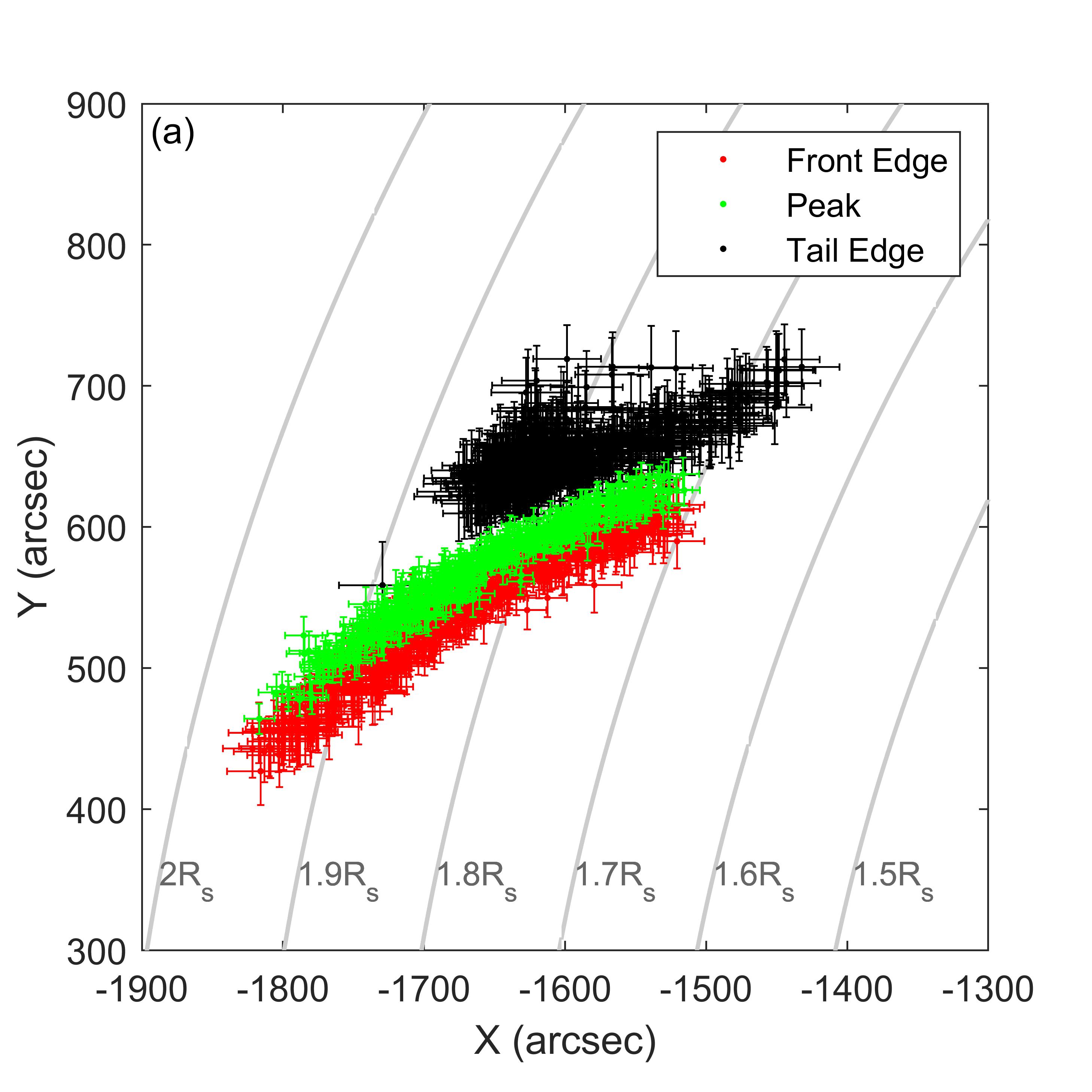}
		\includegraphics[width=0.44\textwidth]{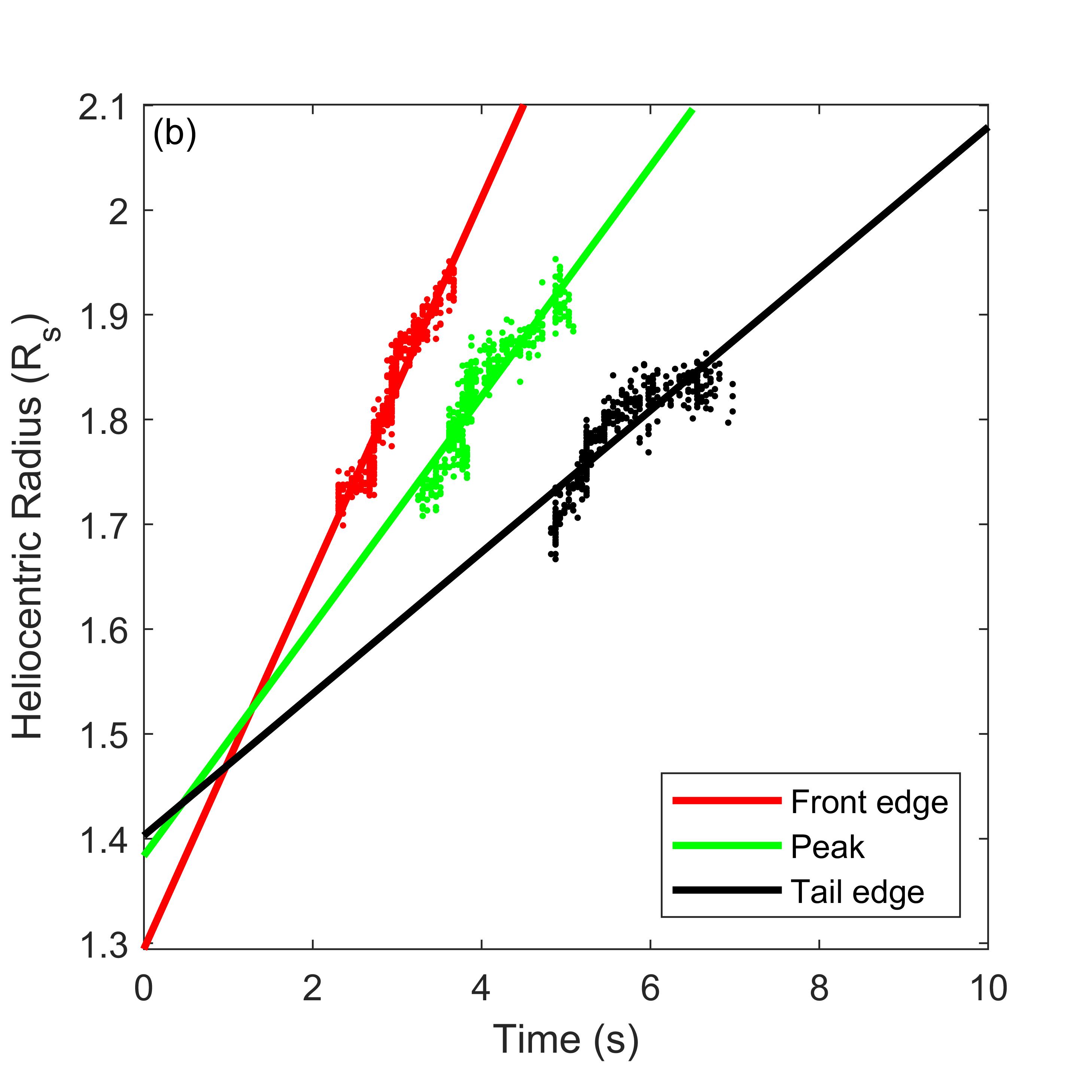}
		\caption{(a) Centroid positions of the front edge, the peak, and the tail edge of the type III radio burst  {in the dynamic spectrum} shown in Figure \ref{fig:1}(c), and the positioning uncertainty (marked as error-bar for each point). (b) Variation of the source height (heliocentric distance) with time (dot points), and the linear least-squares fits (solid lines) of the height-time profile. The results of the front edge, the peak, and the tail edge are presented in the red, green and black color, respectively.
			\label{fig:6}}
	\end{figure}
	
	To quantitatively investigate the electron-velocity dispersion effect on the duration, in the following,  {we will mainly pay attention to the source positions of the front edge, the peak, and the tail edge in the dynamic spectrum, and the radial speed of their electron exciter}. The results are displayed in Figure \ref{fig:6}, where the red, green, and black points represent the leading edge, the peak, and the tail edge, respectively. Panel (a) shows the centroid positions of the radio source in the sky plane, and panel (b) is the height-time profile of the source and its linear least-squares fit. The source height (or heliocentric distance) is derived from the 2-D source center position using $r^2=x^2+y^2$. From Figure \ref{fig:6}, one can see that the source centroids of the front edge, the peak, and the tail edge are well separated from each other spatially. Moreover,  {the electron beams exciting the front edge, the peak, and the tail edge} are moving outward from the Sun with different speed, an average radial velocity of 0.42\,c, 0.25\,c, and 0.16\,c, respectively, based on the linear fit of the source height-time relationship. {Here, it is necessary to point out that the radial speed of the electron beam in the corona may be slightly different from the calculated value, since the observed centroid positions may displace from the actual position due to the ionospheric refraction (e.g. \citealp{gordovskyy2019frequency}). This displacement depends both on the elevation angle of the object and on the wave frequency. For this event, the elevation angle of the Sun is about 54 degree at 12 UT on 6 May 2015. According to the results shown in Figure 2(c) in the paper by \cite{gordovskyy2019frequency}, when the elevation angle is 59 degree, the observed position and the corrected position of Tau A nearly coincide with each other in the frequency range from 30 MHz to 48 MHz. Thus, we do not consider the ionospheric refraction correction for this event, and assume that the relative position of the sources at different frequency is not influenced by ionospheric refraction.}
	
	The above result implies that the electron beam exciter, which is accelerated at the low corona, has distinct velocity dispersion in the radial direction. The radial speed of the non-thermal electrons generating this type III burst may vary in the range from about 0.16\,c to 0.42\,c. Electrons from the flare site with different speed arrive the same altitude at different time. As a result, for a given frequency, the duration of the radio emission will be determined by the time difference when the fastest and the slowest electron beams passing the corresponding source height respectively  {without considering other effects}.

	\begin{figure}[h]
		\centering
		\includegraphics[width=0.81\textwidth]{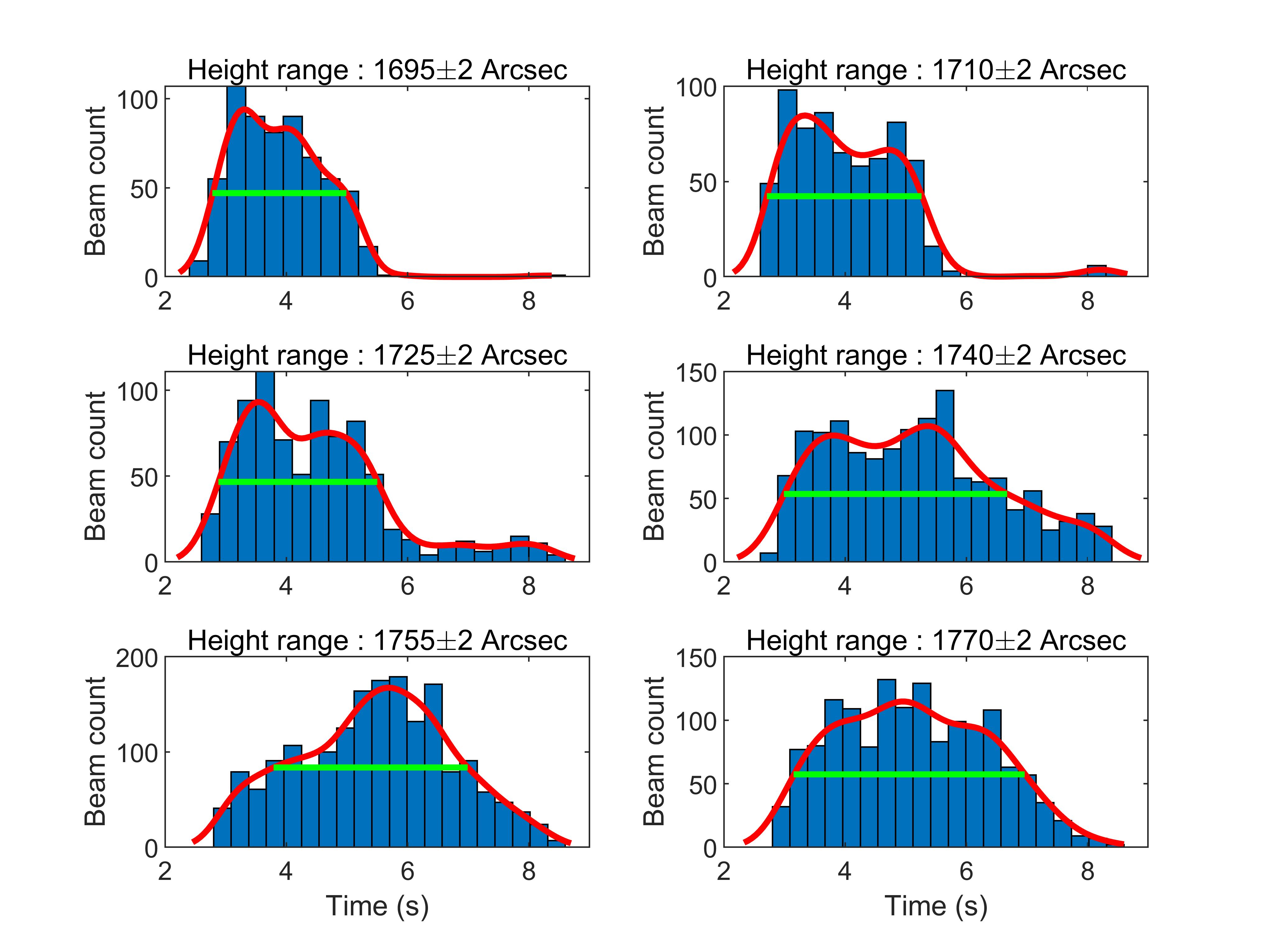}
		\caption{The event number distribution of the arrival time of the sources at six selected heights. The red lines are the smoothed curves using the quadratic regression, and the green lines are the FWHM obtained from the red line. The measured FWHM values ($\Delta t_{\delta v}^\mathrm{FWHM}$) are listed in the the fifth column of Table \ref{tab:1}.
			\label{fig:7}}
	\end{figure}
	
	We can estimate the contribution of the electron-velocity dispersion to the duration from observation by statistical surveying the arrival time of the sources at a given height. Figure \ref{fig:7} shows the event number distribution of the arrival time at six selected heights (within a small range of $\pm 2$  arcsec). We use the FWHM of the smoothed distribution as the contribution of the electron-velocity dispersion to the duration, which is noted as $\Delta t_{\delta v}^\mathrm{FWHM}$. The results are listed in the fifth column of Table \ref{tab:1}. It is found that $\Delta t_{\delta v}^\mathrm{FWHM}$ has larger value for source at higher altitude, namely, the arrival-time difference gets larger with the increase of altitude. The values of $\Delta t_{\delta v}^\mathrm{FWHM}$ are about two to three seconds, which are very close to the values of $\Delta t_{obs}^\mathrm{FWHM}$ in the sixth column measured from the light curve in the frequency channel corresponding to the average  {wave} frequency at the surveying height.
	
	\subsection{Effect of Corona Density Fluctuation}
	The fluctuation of the coronal density can also influence the duration of type III radio burst as proposed by \cite{roelof1989type}. The electron beams may inject into a branch of magnetic field lines. If the radial corona-density profile varies within the branch, for a given wave frequency, the exciting height of the radio wave would be different along different magnetic field lines, as well as the exciting time. According to the result of \cite{roelof1989type}, the duration of type III burst caused solely by corona density fluctuation is
	\begin{equation}
	\Delta t_{\delta N_e}^\mathrm{FWHM} = (0.546\, \mathrm{s})(R_f/R_s)^2(\sigma /\beta_r), \label{eq:8}
	\end{equation}
	where $R_f$ is the heliocentric distance of the source with frequency $f$, $R_s$ is the solar radius, $\sigma=\delta N_e/N_e$  and $N_e$  is the background electron density of corona,  $\delta N_e$ is the variation of the corona density, $\beta_r$ is the radial velocity of the electron beam in the unit of the speed of light.
	
\begin{figure}[h]
		\centering
		\includegraphics[width=0.81\textwidth]{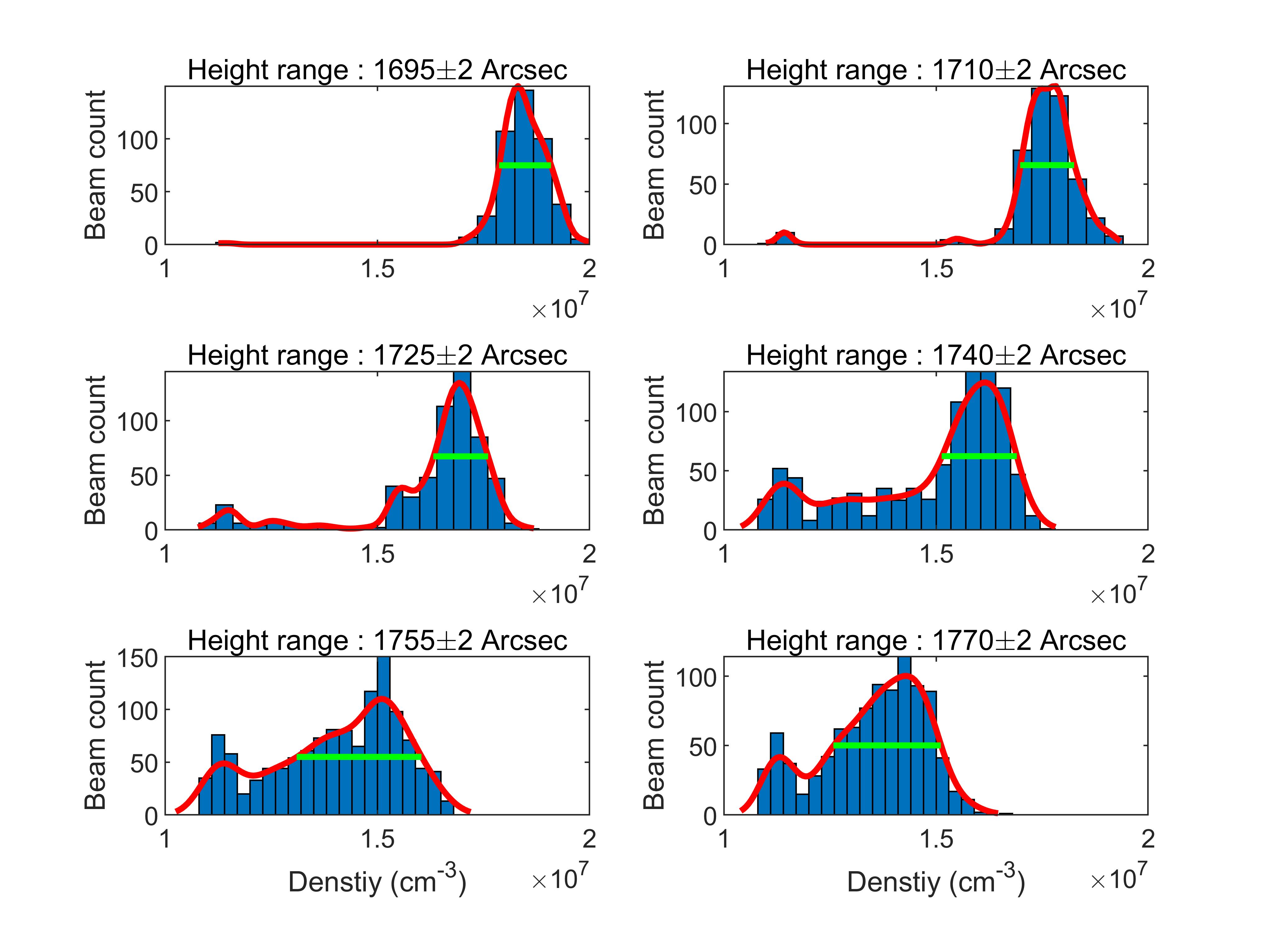}
		\caption{The event number distribution of the electron number density at six selected heights. The red lines are the smoothed curves using the quadratic regression, and the green lines are the FWHM obtained from the red line. \label{fig:8}}
	\end{figure}

	According to the plasma emission mechanism, the radio waves are generated in the local plasma frequency or its harmonic. We can estimate the level of the coronal density fluctuation at a given height by statistically surveying the wave frequency distribution of the sources observed at the same height. Figure \ref{fig:8} shows the corona density distribution at six selected height, where the observed frequency has been converted to the plasma density using the relationship $f_{pe}[\mathrm{kHz}] = 9\sqrt{N_e[\mathrm{cm^{-3}}]}$ between the plasma frequency and the electron density with the fundamental assumption  {(the measured value of $\delta N_e/N_e$ not depending on the fundamental or harmonic assumption)}. The density variation $\delta N_e$ is chosen to be equal to the FWHM of the smoothed distribution. The background density  $N_e$  is given by the average density of all observations at the same height. The third column in Table \ref{tab:1} lists the measured density fluctuation ($\delta N_e/N_e$) at six selected height. It is found that  $\delta N_e/N_e$ generally increases with the increase of altitude  {changing from about 0.07 to 0.19 in the heliocentric distance of about 1.5\,--\,2.0 solar radii}.
	
	Using Equation \ref{eq:8}, we can calculate the duration ($\Delta t_{\delta N_e }^\mathrm{FWHM}$) due to corona density fluctuation. The results are shown in Column 4 in Table \ref{tab:1}. In the calculation, $\beta_r$ is chosen to be 0.25\,c, namely, the radial velocity of the radio source of the peak. It is found that the duration ($\Delta t_{\delta N_e }^\mathrm{FWHM}$) increases with the increase of source altitude or with the decrease of wave frequency. The observed duration ($\Delta t_{obs }^\mathrm{FWHM}$) is about 2.2 to 5.7 times of duration ($\Delta t_{\delta N_e }^\mathrm{FWHM}$) caused by density fluctuation.  {The contribution of density fluctuation on the duration is below the expectation of \cite{roelof1989type}, since the measured level of density fluctuation is less than the value of $\delta N_e/N_e=0.3$ they chose.}
	
	\begin{table}[]
		\caption{The durations of the type III burst  {observed by LOFAR at a given frequency (observation at multiple frequencies between 34 and 38 MHz are presented)}, and the contributions from electron-velocity dispersion and coronal density fluctuation.  {The first and second columns are the source height for surveying and the average frequency of waves observed at the corresponding height. The third column is the density-fluctuation level measured at the height}. The fourth and fifth columns are the duration contributions from the density fluctuation and the electron-velocity dispersion, respectively. The sixth column is observed duration in the frequency channel with the frequency in Column 2. \label{tab:1}}
		
		\centering
		\begin{tabular}{cccccc}
			\hline
			$R$ (arcsec) & $\bar{f}$ (MHz) & $\delta N_e /N_e$ & $\Delta t_{\delta N_e}^{\rm FWHM}$(s) & $\Delta t_{\delta v}^{\rm FWHM}$ (s) & $\Delta t^\mathrm{FWHM}_{obs}$ (s) \\ \hline
			1695$\pm$2 & 38.51$\pm$0.07 & 0.067 & 0.44 & 2.21 & 2.52 \\
			1710$\pm$2 & 37.96$\pm$0.12 & 0.071 & 0.48 & 2.55 & 2.52 \\
			1725$\pm$2 & 37.03$\pm$0.15 & 0.076 & 0.52 & 2.62 & 2.73 \\
			1740$\pm$2 & 36.17$\pm$0.13 & 0.110 & 0.77 & 3.68 & 2.78 \\
			1755$\pm$2 & 34.98$\pm$0.10 & 0.195 & 1.39 & 3.20 & 3.09 \\
			1770$\pm$2 & 34.01$\pm$0.09 & 0.177 & 1.29 & 3.81 & 3.46 \\ \hline
		\end{tabular}
	\end{table}

	\subsection{Effect of Wave Propagation}
	As mentioned in the introduction, wave scattering and refraction can also increase the observed duration by varying the light-path length of the waves (e.g. \citealp{steinberg1971coronal}; \citealp{arzner1999radiowave}). Since the intrinsic source property at the wave excitation site is unknown from remote sensing, it is difficult to analyze the contribution of the wave propagation effect on the duration directly from observation.
	
\begin{figure}[h]
		\centering
		\includegraphics[height=0.3\textheight]{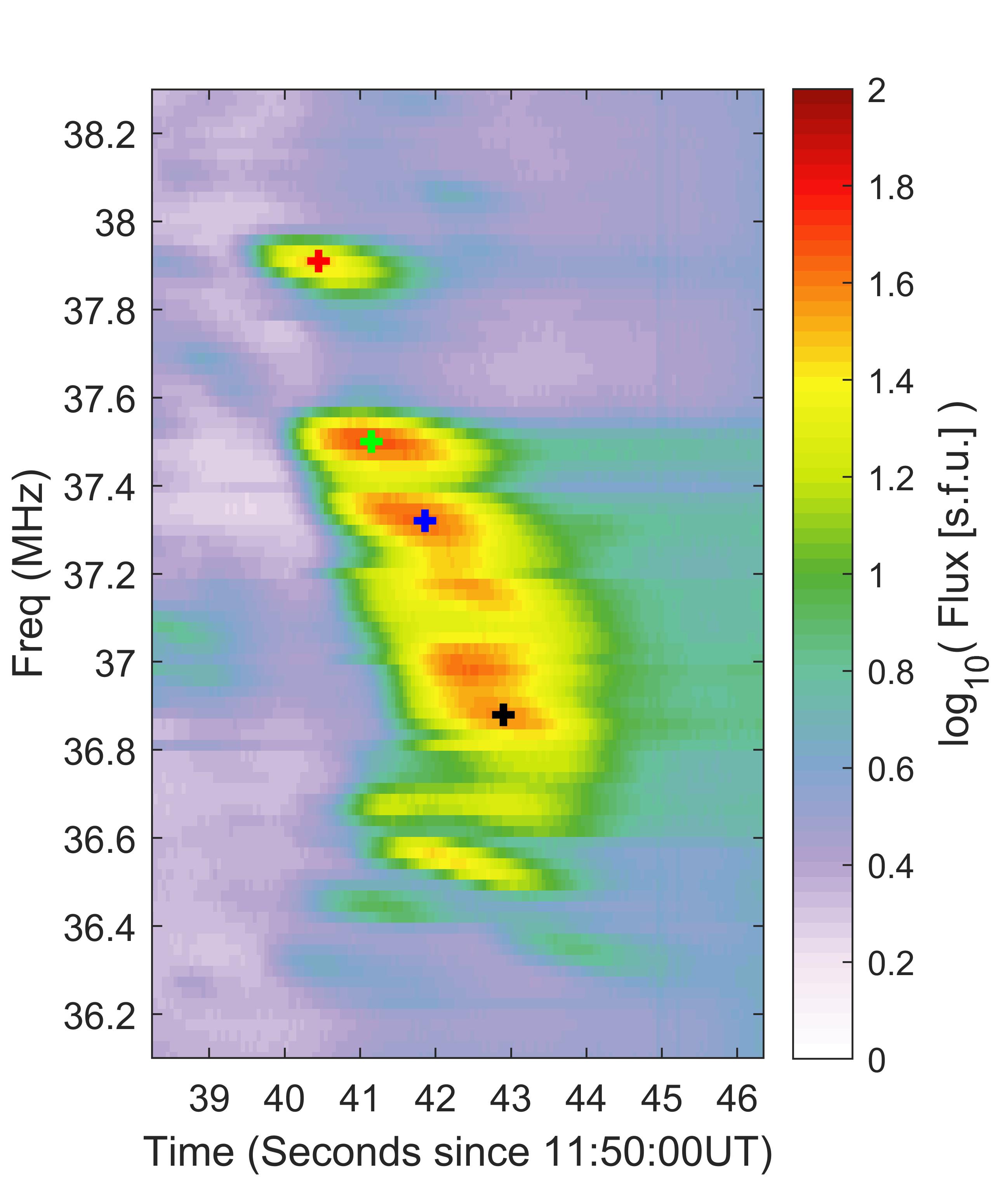}
		\includegraphics[height=0.3\textheight]{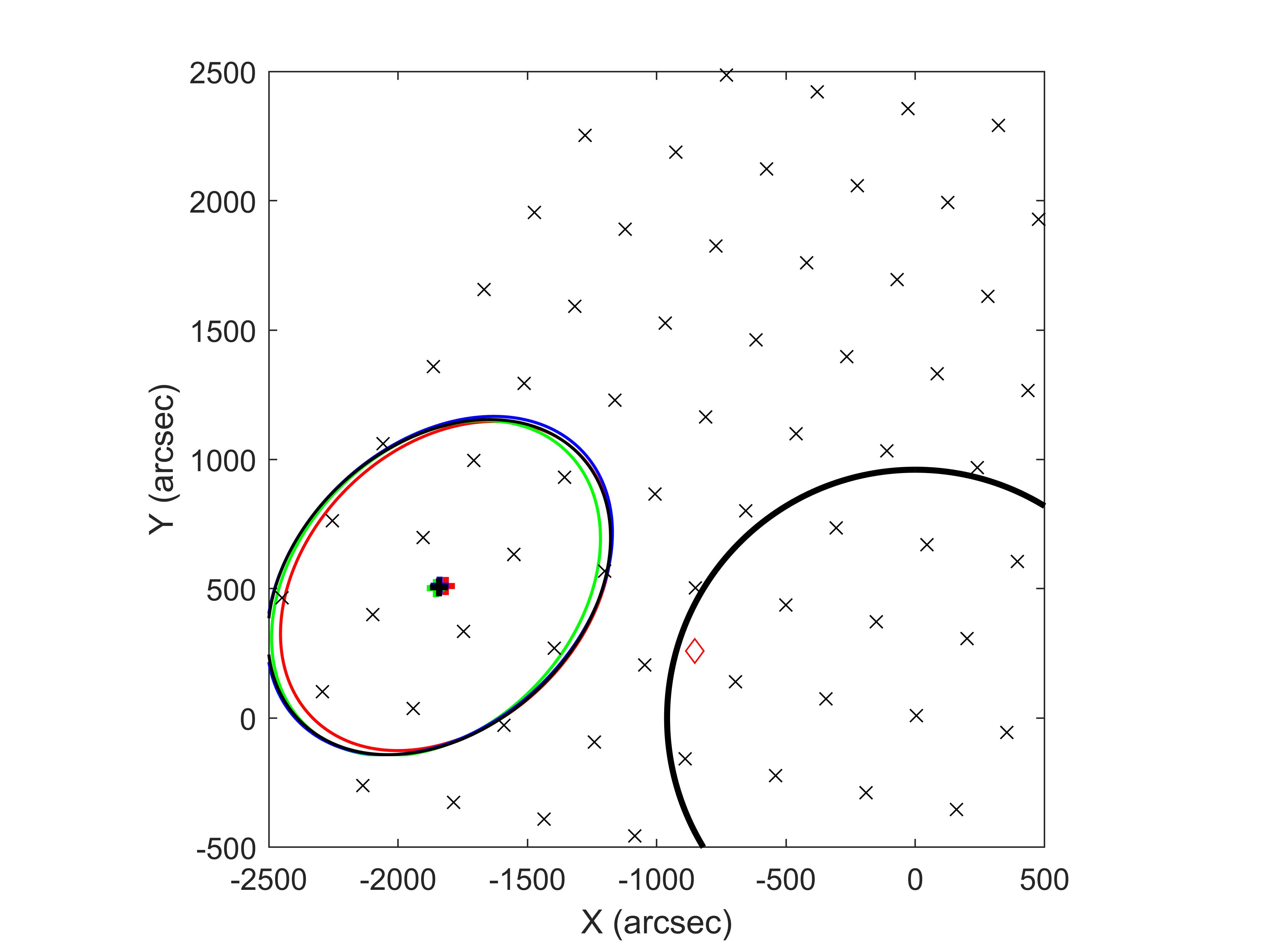}
		\caption{The enlarged dynamic spectrum (left panel) of the short-term burst, and the source centroid position and size (right panel) of the waves marked by the color “plus” on the left panel.	\label{fig:9}}
	\end{figure}

	Fortunately, there are a few shorter-term radio bursts just before this type III burst, as shown in Figure \ref{fig:1}(b). These shorter-term bursts took place at about 11:50:40\,UT (50 seconds before this type III radio burst), with peak flux that is comparable to the intensity of the type III radio burst we study. The enlarged dynamic spectrum of the shorter-term bursts is shown on the left panel in Figure \ref{fig:9}, while the right panel displays the centroid position and elliptical-Gaussian fitted source size of four radio sources. The wave frequency and observation time of the four sources are marked by four “plus” with different color in the spectrum on the left panel.
	
	One can see that the source of the shorter-term burst locates nearly at the same position of the type III burst in the sky plane. Thus, it is reasonable to assume that radio waves from the shorter-term burst and the type III burst with the same frequency had experienced similar scattering and refraction processes during their propagation in the solar atmosphere. The observed duration of shorter-term bursts can be set as an upper limit of the contribution of the wave propagation effect to the duration.

	Applying the asymmetric exponential function fit to the light curve, the durations of shorter-term burst are found to be 1.26\,s, 1.68\,s, 1.73\,s, and 1.88\,s in the frequency channels of 37.91\,MHz, 37.49\,MHz, 37.32\,MHz and 36.88\,MHz, which are marked by the red, green, blue, and black “plus” on the left panel in Figure \ref{fig:9}, respectively. Considering the duration of the type III radio burst near the frequency of 38\,MHz is about 2.52\,s, the upper limit contribution of the wave propagation effect is about half of the observed duration. In fact, the waves of the shorter-term burst are not likely to come from an ideally impulsive point source, but having an intrinsic growth-damping time and space extension, so the contribution of wave propagation effect should be less than half of the observed duration. 
	
	\section{Summary}
	
	In this work, the source position and duration of a limb type III solar radio burst at 11:51:30\, UT on 6 May 2015 in the frequency range of 30\,--\,41\,MHz are measured, using the spectrum and the tied-array beam-formed images observed by LOFAR with high temporal and frequency resolution. It was a single and clear type III radio burst with duration of about 3 seconds. After getting the centroid positions of the radio waves at different time and with different frequency, we linearly fitted the radial velocity of the source movement, and statistically surveyed the source arrival time and the fluctuation level of corona density at several selected altitude. Consequently, we have evaluate the contribution of three factors on the observed duration of type III burst, namely, the electron-velocity dispersion effect, the corona density fluctuation effect, and the wave propagation effect by comparison with a few shorter-term bursts from the same region. The main results are summarized in the following.
	
	\begin{itemize}
		\item 	The  {source} centroid positions of the front edge, the peak, and the tail edge in the dynamic spectrum are spatially separated in the sky plane as shown in Figure \ref{fig:6}(a). This indicates that they are generated by electron beams moving in different magnetic flux tube  {in the corona}. Their  {electron exciters} move outward from the Sun with radial velocity of about 0.42\,c, 0.25\,c, and 0.16\,c, respectively. The duration caused by this velocity dispersion is the dominant contribution on the observed duration of the type III burst.
		\item  	The electron-density fluctuation   in the corona increases with height, changing from about 0.07 to 0.19 in the heliocentric distance of 1.5\,--\,2.0\,solar radii. The observed duration of the type III burst is about 2.2 to 5.7 times of the duration caused by density fluctuation.
		\item  {The shorter-term bursts took place just at about fifty seconds before the type III burst. If we assume the radio waves of the short burst and the type III burst experience similar scattering during the propagation, we can infer that the contribution of wave propagation effect on the duration of this type III burst is no more than half of the observed duration.}	
	\end{itemize}
	
	Figure \ref{fig:6}(a) shows that waves on the front edge  { of the dynamic spectrum} are excited at higher altitude, comparing with the waves on the tail edge with the same frequency. This means that the fast electrons producing the front edge propagate outward in a magnetic flux tube with  {high} corona density, while the slow electrons producing the tail edge move in a flux tube with  {low} corona density.  {Let $ h_1 $ and $ h_2 \, (<h_1) $ represent the source height of the front edge and the tail edge at the same frequency $ f $,  $ v_1 $ and $ v_2 \, (<v_1) $ are the radial speed of the electron beams exciting the front edge and the tail edge in the dynamic spectrum, respectively. Here, the height starts from the electron acceleration site.  The duration due to the exciter-velocity difference and the variation of background density can be simplified approximately as 
		\begin{equation}
		\Delta t = \dfrac{h_2}{v_2} - \dfrac{h_1}{v_1} =\dfrac{(v_1-v_2)}{v_1}\dfrac{h_2}{v_2}-\dfrac{(h_1-h_2)}{v_1}.
		\label{eq:9}
		\end{equation}
		The second term on the right side is the contribution of density variation on the duration, which is negative for $ h_1>h_2 $.} Thus, for this type III event, the corona density fluctuation effect  {(in the sense of fibrous corona of \cite{roelof1989type})} has a negative contribution to the observed duration, in other words, $\Delta t^\mathrm{FWHM}_{obs}=\Delta t_{\delta v}^{\rm FWHM}-\Delta t_{\delta N_e}^{\rm FWHM}+\delta t$ , where  {$\delta t$ is the duration caused by the wave propagation effect  (including radio wave scattering between the source and the observer) and the intrinsic generation process of the EM wave}. This is different from the work of \cite{roelof1989type} in which the non-thermal electrons are all assumed to move outward with the same velocity.
	
	{The duration contribution from the intrinsic radio emission process is not investigated in this study. In principle, it is difficult to separate it from the wave propagation effect without \textit{in situ} observation of the source property at the wave excitation site in the corona.  In the future, we may firstly calculate the  contribution of wave propagation effect by numerical simulation such as ray tracing, using the measured density-fluctuation level from LOFAR observation. Then, the contribution of the intrinsic emission process can be obtained by subtracting the contributions of other effects from the observed duration.}

	\begin{acknowledgements}
		This paper is based in part on data obtained from facilities of the International LOFAR Telescope (ILT) under project code L342370. LOFAR is the LOw Frequency ARray designed and constructed by ASTRON.  {LOFAR has observing, data processing, and data storage facilities in several countries, that are owned by various parties (each with their own funding sources), and that are collectively operated by the ILT foundation under a joint scientific policy. The ILT resources have benefitted from the following recent major funding sources: CNRS-INSU, Observatoire de Paris and Universite d'Orleans, France; BMBF, MIWF-NRW, MPG, Germany; Science Foundation Ireland (SFI), Department of Business, Enterprise and Innovation (DBEI), Ireland; NWO, The Netherlands; The Science and Technology Facilities Council, UK; Ministry of Science and Higher Education, Poland.} The authors would also thank the science team of AIA and RHESSI, which made the EUV and X-Ray image data available. AIA data were supplied courtesy of the SDO/AIA consortia (contract NNG04EA00C of the SDO/AIA instrument to LMSAL). The RHESSI satellite is a NASA Small Explorer (SMEX) mission. The research at USTC was supported by the National Nature Science Foundation of China (Grant No.41574167 and 41174123) and the Fundamental Research Funds for the Central Universities.  {S.Y. are supported by NASA grant NNX17AB82G and NSF grants AGS-1654382 and AST-1735405 awarded to the New Jersey Institute of Technology.  E.P.K. was supported from the Science and Technology Facilities Council  (STFC)  Consolidated  Grant  ST/P000533/1. }
	\end{acknowledgements}	
	\bibliography{cite}
	
	\clearpage

\end{document}